\documentclass[12pt]{iopart}
\pdfoutput=1
\expandafter\let\csname equation*\endcsname\relax
\expandafter\let\csname endequation*\endcsname\relax
\usepackage[pdftex]{graphicx,color}
\usepackage{amsmath,amssymb,wasysym,graphicx,capt-of,ifthen,calc}
\usepackage{enumerate}
\usepackage{url}
\usepackage{subfigure}
\usepackage{float}
\begin{document}

\title{Fixation in Fluctuating Populations}
\author{Deepak Bhat$^*$, Jordi Pi\~nero\dag\ddag, and S. Redner$^*$}
\address{$^*$Santa Fe Institute, 1399 Hyde Park Road, Santa Fe, NM, 87501 USA}
\address{\dag Complex Systems Lab, ICREA-Universitat Pompeu Fabra, Dr Aiguader 88, 08003
  Barcelona, Spain}
\address{\ddag Institut de Biologia Evolutiva, Psg.\ Barceloneta 37-49, 08003 Barcelona, Spain}

\begin{abstract}
  We investigate the dynamics of the voter model in which the population
  itself changes endogenously via the birth-death process.  There are two
  species of voters, labeled A and B, and the population of each species can
  grow or shrink by the birth-death process at equal rates $b$.  Individuals
  of opposite species also undergo voter model dynamics in which an AB pair
  can equiprobably become AA or BB with rate $v$---neutral evolution.  In the
  limit $b/v\to\infty$, the distribution of consensus times varies as
  $t^{-3}$ and the probability that the population size equals $n$ at the
  moment of consensus varies as $n^{-3}$.  As the birth/death rate $b$ is
  increased, fixation occurs more more quickly; that is, population
  fluctuations promote consensus.
\end{abstract}
\maketitle

\section{INTRODUCTION}

A fundamental concept in evolutionary dynamics is that of \emph{fixation}.
In a population that consists of two (or more) species, demographic
fluctuations or competitive effects can lead to a long-time state in which
only one species remains, or fixates~\cite{M62,K83,E04,N06,CK09}.  This
fixation process has been extensively investigated in situations where the
dynamics is defined to keep the total population constant.  Indeed, in many
evolutionary dynamics experiments on controllable systems, such as bacterial
colonies, a typical protocol is to cull the population at fixed time
intervals so that the population is the same at each of these resetting
events~\cite{GMB17}.  However, in real bacterial colonies, the number of
organisms changes with time.  A pertinent example is when each species
undergoes birth-death dynamics with equal birth and death rates for each
species so that the average population is fixed but fluctuates endogenously.

\begin{figure}[ht]
  \centerline{ \includegraphics[width=0.4\textwidth]{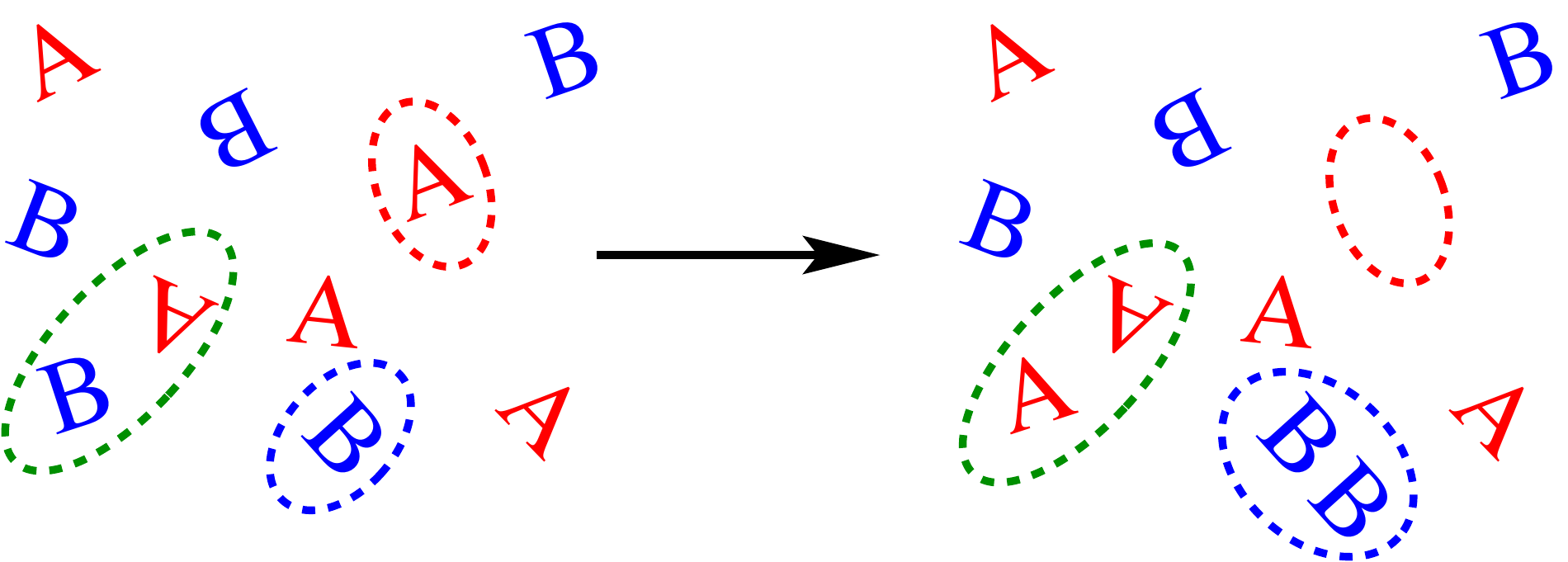} }
  \caption{\small Cartoon of update events in the fluctuating voter model.
    An A dies with rate $\lambda$ (red dashed oval), a B gives birth to
    another B also with rate $\lambda$ (blue oval), and a circled AB pair
    changes to AA with rate $v$ (green oval).}
\label{cartoon}  
\end{figure}

We introduce the \emph{fluctuating voter model} (FVM) to understand fixation
in such a population.  Here, two distinct species of voters can change their
opinion state by voter-model dynamics~\cite{KRB10,L99,K92} \emph{and} the
populations can grow or shrink by birth and death~\cite{K49,KRB10}
(Fig.~\ref{cartoon}).  Our perspective is complementary to the modeling of
biological populations in randomly switching
environments~\cite{HLGK16,HSM17,WFM17}.  The two species are equivalent in
all respects except their identity.  In a voter model update, which occurs at
rate $v$, an AB pair transforms equiprobably to either AA or BB; that is, the
evolution is neutral.  We investigate the perfectly mixed limit, in which any
pair of opposite-opinion voters is equally likely to interact.  The voter
model update is repeated \emph{ad infinitum} or until fixation (consensus) is
reached, where only a single species remains.  In addition, each individual
can give birth to an offspring of the same type as the parent at rate
$\lambda$, and each individual can die with rate $\mu$.

By these mechanisms, the population and its composition change with time.
Except for the pathological situation where the population grows
exponentially in time (see, e.g., Ref.~\cite{MR14}), consensus is eventually
reached.  We assume that the birth and death rates equal a common value,
$\lambda=\mu\equiv b$, so that the average population is fixed, but
population fluctuations grow with time.  Our main results are: (a) For
$v/b \to 0$, fixation necessarily occurs and the distribution of fixation
times $F(t)$ scales as $t^{-3}$; at fixation, the probability $Q_n$ that the
population size equals $n$ scales as $n^{-3}$.  (b) Population fluctuations
promote fixation; the fixation time is a decreasing function of the
birth/death rate $b$.  (c) For arbitrary $b$ and $v$, the fixation time
distribution $F(t)\sim t^{-1-\beta}$, with $\beta$ a function of $b$ and $v$.

In the next section, we first treat the limit of $v/b\to 0$, where the system
reduces to two uncoupled birth-death processes.  Although the dynamics of a
single birth-death process is very well understood, the properties of
multiple birth-death processes appears unexplored, and we determine many of
its basic properties analytically (see also the appendices).  In
Sec.~\ref{sec:v>0}, we then outline our main results for the FVM.

\section{UNCOUPLED LIMIT}

In the limit $v/b\to 0$, voter model updates do not occur and the birth-death
processes for the two species decouple.  We may therefore apply well-known
results for the birth-death process to infer the extinction dynamics.  For
the single-particle initial condition, with the birth and death rates set to
a common value $b$, the probability that there are $n$ particles at time $t$
is (see also Appendix~\ref{sec:gf})~\cite{K49,KRB10}
\begin{align}
  \label{Pnt}
P_n(t) =\frac{(b t)^{n-1}}{(1+b t)^{n+1}}\qquad\qquad  P_0(t)=
  \frac{b t}{1+b t} \,,
\end{align}
from which the average population is $\langle n(t) \rangle=1$, while the
variance $\sigma^2\equiv\langle n(t)^2\rangle-\langle n(t)\rangle^2=2b t$.
Thus even though $\langle n(t) \rangle=1$, there are huge population
fluctuations between different realizations of the birth-death process.

Although the average population is fixed, its ultimate fate is extinction.
From the second of Eqs.~\eqref{Pnt}, the survival probability $S_1(t)$,
namely, the probability that a single birth-death process does not go extinct
by time $t$ is
\begin{subequations}
  \label{SF}
\begin{align}
  \label{S1}
  S_1(t)=1-P_0(t)=\frac{1}{1+bt}\,,
\end{align}
while the probability that extinction occurs at time $t$ is
\begin{align}
  \label{F1}
  F_1(t) = -\frac{d }{dt} S_1(t) = \frac{b}{(1+bt)^2}\,.
\end{align}
\end{subequations}
This birth-death process is \emph{recurrent} (analogous to diffusion in one
dimension~\cite{F68,R01}), because the extinction probability $P_0\to 1$ for
$t\to \infty$, but the average time to reach extinction,
$\langle t\rangle = \int_0^\infty dt\, t\, F_1(t)$ is infinite.

\subsection{Two  Identical Birth-Death Processes}
\label{subsec:2}

\emph{Initial state: (A,B) = (1,1)}: We study the dynamics of two uncoupled
birth-death processes in which the initial state consists of one $A$ and one
$B$, and the common birth/death rates of each process are the same and equal
to $b$.  For extinction to not occur by time $t$, the number of particles in
both of the two birth-death processes must remain non-zero.  This probability
is
\begin{subequations}
\begin{align}
    \label{S2}
  S_2(t)= [S_1(t)]^2=[1-P_0(t)]^2=\frac{1}{(1+b t)^2}\,.
\end{align}
This quantity is also the probability that the extinction time is $t$ or
greater.  Thus the probability that one of the two species goes extinct at
time $t$ is
\begin{align}
  F_2(t)= -\frac{d}{dt} [S_2(t)]= \frac{2b}{(1+b t)^3}\,.
\end{align}
\end{subequations}
Because the exponent of this time dependence is less than $-2$, the average
extinction time is finite:
\begin{align}
  \label{tav}
  \langle t\rangle = \int_0^\infty dt\, t\, F_2(t)
       = \int_0^\infty dt\, S_2(t) = \frac{1}{b}\,.
\end{align}
In contrast to a single birth-death process, the \emph{smallest} extinction
time among two independent birth-death processes is finite.  A related
dichotomy occurs in one-dimensional diffusion~\cite{R01}: the average time
for a single diffusing particle that starts at $x$ to reach $x=0$ is
infinite, but for three particles that start at $x$, the smallest time for
one of them to reach $x=0$ is finite.  Even though the average extinction
time for two birth-death processes is finite, the mean-square time extinction
time is divergent.  Thus in a finite number of realizations of two
independent birth-death processes, there will be huge sample-to-sample
fluctuations in the time when the first species goes extinct.

At extinction, a natural characteristic is the average number of particles
$\langle n\rangle$ of the surviving species.  Since the birth-death process
conserves the average particle number and the initial state consists of 2
particles, there must be $2$ particles, on average, at any time, including
the moment when one species goes extinct.  We may also determine $Q_n$, the
probability distribution for $n$.  At time $t$, the probability that the
number of particles in either species equals $n$ is given by $P_n(t)$ in
Eq.~\eqref{Pnt}.  To obtain $Q_n$, we convolve this distribution with the
probability that the other birth-death process goes extinct at time $t$,
namely $F_1(t)$ in Eq.~\eqref{F1}.  Thus the probability that the surviving
population consists of $n$ particles when the first species goes extinct is
\begin{align}
\label{Qn}
  Q_n = 2\int_0^\infty dt\, F_1(t) P_n(t)  = 2\int_0^\infty dt\,\,
  \frac{b\,(b t)^{n-1}}{(1+b t)^{n+3}}
  =   \frac{4\Gamma(n)}{\Gamma(n+3)}\simeq \frac{4}{n^3}\,.
\end{align}
The prefactor 2 accounts for the fact that either of the two species could go
extinct first.  Fortuitously, this expression for $Q_n$ is identical to the
degree distribution in linear preferential attachment networks~\cite{KR01}.
We do not have any explanation for this remarkable coincidence.  It is also
straightforward to verify that when one species goes extinct the number of
particles of the remaining species is
$\langle n\rangle = \sum_{n\geq 1} nQ_n =2$.  \smallskip

\emph{Initial state: (A,B) = $(k,k)$}: We now briefly study the initial state
with $k>1$ particles of each species in the initial state.  The expression
for $P_n(t)$ for $n>0$ becomes more unwieldy as $k$ increases, and we only
investigate the extinction dynamics.  The probability that a single
birth-death process with $k$ particles in the initial state goes extinct at
time $t$ is $P_0(t) = \left[{bt}/(1+bt)\right]^k$ (see
Appendix~\ref{sec:gf}).  From this expression, the probability that this
birth-death process survives until time $t$ is $S^{(k)}\!(t)= 1-P_0$.  The
probability that two independent birth-death processes with the $(k,k)$
initial condition survive until time $t$ is $\big[S^{(k)}\!(t)\big]^2$.
Consequently, the average time at which one of the two species first goes
extinct is
\begin{align}
  \label{t-kk}
  \langle t\rangle &= \int_0^\infty dt\, \big[S^{(k)}\!(t)\big]^2
  = \int_0^\infty dt\,\left[ \frac{(1+bt)^k-(bt)^k}{(1+bt)^k}\right]^2\nonumber\\[3mm]
                   &= \frac{2k}{b}\big[H_{2k}-H_k\big]\simeq \frac{2\,k \ln 2}{b}\,,
\end{align}
where $H_n$ is the $n^{\rm th}$ harmonic number,
$H_n=\sum_{1\leq k\leq n}\frac{1}{k}$.  The integral was performed using
Mathematica~\cite{W}, but an important preliminary step to get a simple
result is to make the substitution $y=1/bt$ in the integrand.
  
\section{The Fluctuating Voter Model (FVM)}
\label{sec:v>0}

\begin{figure}[ht]
\centerline{ \includegraphics[width=0.9\textwidth]{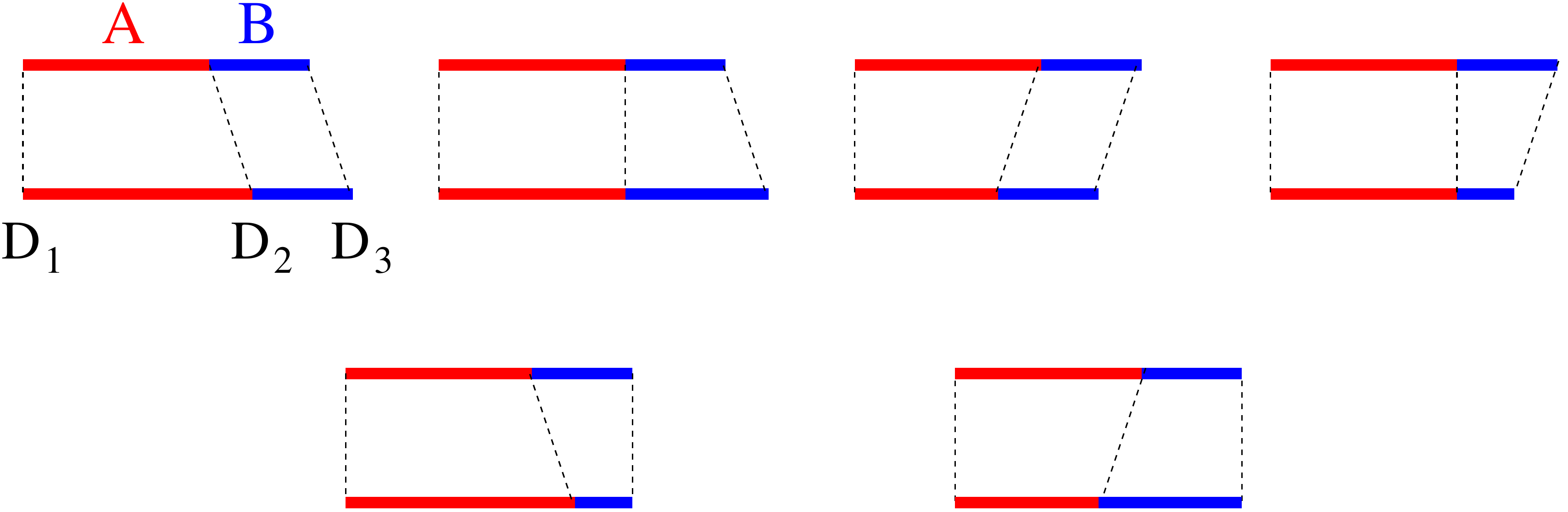} }
\caption{\small Representation of the FVM as an interval of length $N_A+N_B$.
  Top row: changes due to birth of an A, birth of a B, death of an A, and
  death of a B, respectively.  Bottom row: changes due to the voting events
  AB $\to$ AA and AB $\to$ BB, respectively.  The diffusion coefficients of
  the effective interface particles are indicated. }
\label{fvm}  
\end{figure}

We now investigate the dynamics of the FVM when the voting rate $v$ and the
birth/death rate $b$ are both nonzero.  Let $N_A$ and $N_B$ denote the
respective number of voters of type $A$ and $B$ in a population of
$N=N_A+N_B$ individuals, and let $x=N_A/N$ and $1-x=N_B/N$ be the fraction of
voters in each state.  When the system is perfectly mixed, the total rate for
an event to occur (either voting or birth/death) for $N\gg 1$ is
$R=2vNx(1-x)+2b N$.  We take the voting rate $v=1$ henceforth and study the
dynamics as a function of the birth/death rate $b$.  With probability
$2Nx(1-x)/R$ a voting event occurs in which an AB pair changes equiprobably
to AA or BB.  With the complementary probability $2b N/R$, an individual
either gives birth or dies (Fig.~\ref{fvm}).  After each update, the time is
incremented by an exponential random variable with mean value $1/R$.  These
updates are repeated until consensus is reached.  The above defines the
event-driven algorithm~\cite{G77} for the time evolution.

To understand the dynamics of FVM, it is helpful to first map its dynamics
onto that of the reunion of three diffusing particles on the line, and
finally to exploit known results about this three-particle problem~\cite{R01}
to predict the survival time distribution exponent.  We first represent the
FVM as a line interval that consists of two subintervals of lengths $N_A$ and
$N_B$ (Fig.~\ref{fvm}).  The events of birth, death, and voting lead to the
changes in the interval lengths indicated in the figure.  Whenever the
boundary between A's and B's reaches either the left or right end of the
interval, extinction of one species occurs.  The left end of the interval is
stationary, by construction, and thus has diffusion coefficient $D_1=0$.  By
examining Fig.~\ref{fvm}, we deduce that the AB interface particle and the
right edge of the interval have respective diffusion coefficients
\begin{align}
  \label{D}
  D_2 = \frac{2Nx(1-x)+2bNx}{R} \qquad\qquad  D_3 = \frac{2bN}{R}\,.
\end{align}

This 3-particle system, with particles located at $(x_1,x_2,x_3)$, can be
mapped onto the diffusion of a single particle at $(x_1,x_2,x_3)$ in three
dimensions with absorbing boundary conditions whenever $x_1=x_2$ or
$x_2=x_3$.  This corresponds to the middle particle of the 3-particle system
reaching either end of the interval.  In turn, this effective single-particle
system in three dimensions subject to the constraint that the walk dies
whenever $x_1=x_2$ or $x_2=x_3$ is isomorphic to a single diffusing particle
in a two-dimensional absorbing wedge of opening angle $\theta$, with (see
Ref.~\cite{R01} for a detailed explanation of this geometric argument)
\begin{subequations}
\begin{align}
  \label{theta}
  \theta = \cos^{-1}\left[\frac{D_2}{\sqrt{(D_1+D_2)(D_2+D_3)}}\right]\,.
\end{align}
The survival probability for the middle particle, which is the same as the
probability that extinction has not yet occurred is known to scale as
$t^{-\beta}$, with $\beta=\pi/2\theta$~\cite{R01}.  Figure~\ref{beta-vs-b}(a)
shows our simulation results for the time dependence of the survival
probability, which indicates a power-law temporal decay with a non-universal
exponent.

\begin{figure}[ht]
  \centerline{
\subfigure[]{\qquad\includegraphics[width=0.47\textwidth]{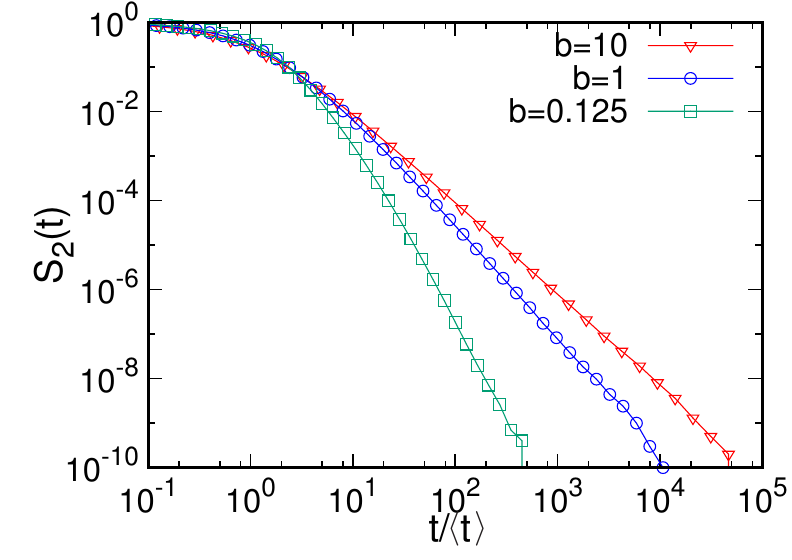}}\qquad\qquad
  \subfigure[]{\includegraphics[width=0.49\textwidth]{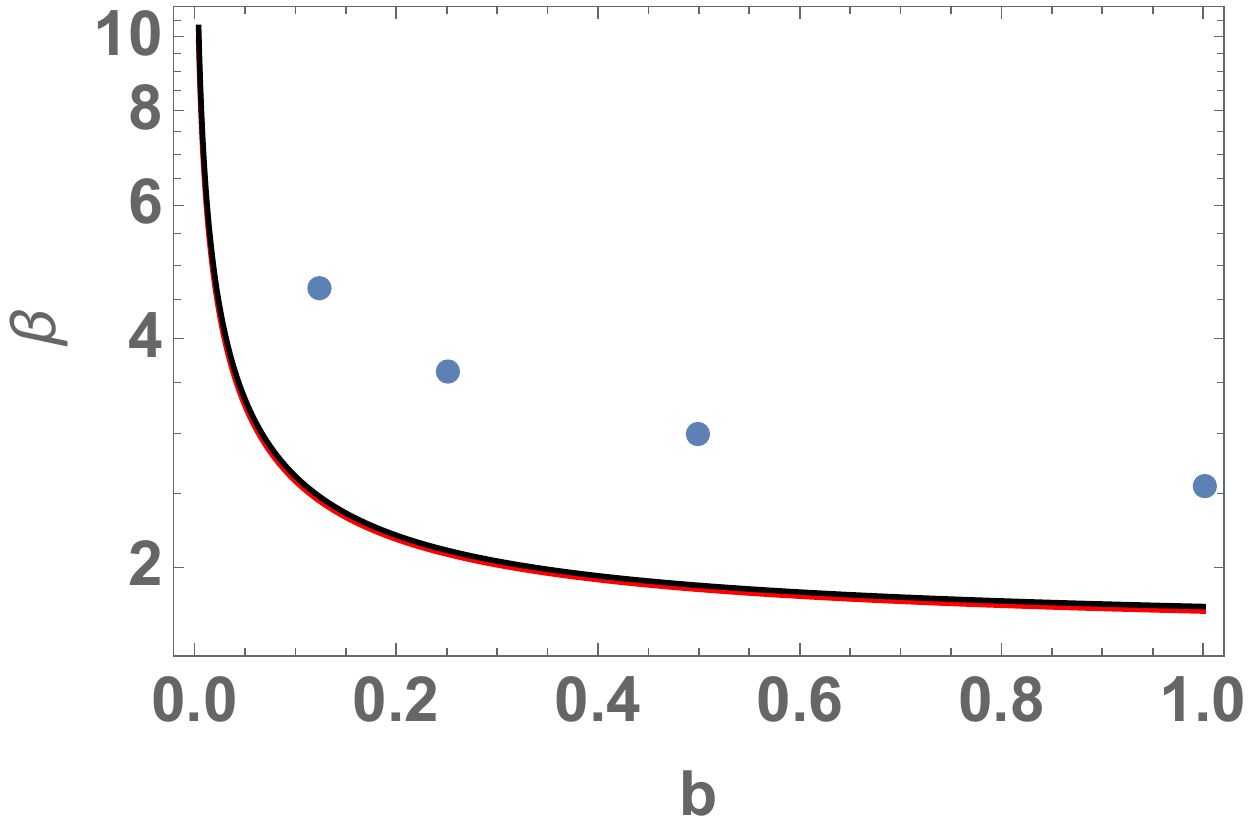}}}
\caption{\small (a) Survival probability $S_2(t)$ versus $t/\langle t\rangle$
  for representative $b$ values.  The data are based on $10^{10}$
  realizations.  (b) The survival probability exponent $\beta$ versus birth
  rate $b$ predicted by~\eqref{beta} (black) and the more principled
  averaging procedure discussed in the text (red).  The points are estimates
  for $\beta$ from simulations.}
\label{beta-vs-b}  
\end{figure}

To determine the exponent $\beta$, we need to apply Eq.~\eqref{theta} to the
FVM.  Here we need to account for the position ($x$) dependence of the
diffusion coefficients in Eq.~\eqref{D}.  The simplest scheme is to merely
replace the true diffusion coefficients in \eqref{D} with their values when
the expressions $x(1-x)$ and $x$ in \eqref{D} are averaged over the interval.
For this prescription, we assume that the interface position is uniformly
distributed over the interval, which gives
$\langle x(1-x)\rangle=\frac{1}{6}$.  In fact, the probability distribution
of $x$ is uniformly distributed over the interval for the mean-field voter
model in the long-time limit~\cite{AC90}.  With this ansatz, we obtain, after
some simple algebra,
\begin{equation}
  \label{beta}
  \beta = {\pi}\Big/{2 \cos^{-1}\left[\frac{1+3b}{\sqrt{1+12b+27b^2}}\right]}\,.
\end{equation}
\end{subequations}
A more principled procedure would be to include the $x$-dependence of the
diffusion coefficients in the expression for $\beta=\pi/\big[2\theta(x)\big]$
and then numerically average this expression uniformly over the interval.
This procedure leads to a result that closely matches \eqref{beta}
(Fig.~\ref{beta-vs-b}(b)).

The main features of Eq.~\eqref{beta} is that the exponent $\beta$
monotonically increases as $b$ decreases (and is slowly varying in $b$ for
$b \gtrsim\frac{1}{2}$).  The values of the survival probability exponent
$\beta$ from Eq.~\eqref{beta} as a function of $b$ is shown in
Fig.~\ref{beta-vs-b}(b).  Estimates of $\beta$ from simulation data for
various $b$ values are also shown in this figure to give a sense of the
accuracy of our analytical approach.  It is not feasible to obtain reliable
estimates of $\beta$ from simulation for smaller $b$ because the exponent
becomes quite large.  Conversely, for larger $b$, the exponent $\beta$ from
simulations is nearly constant.

\begin{figure}[ht]
  \centerline{\includegraphics[width=0.45\textwidth]{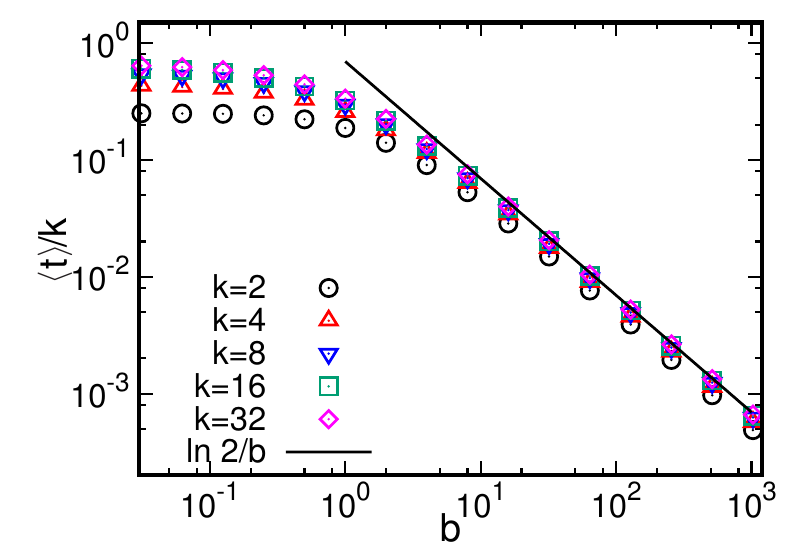}}
  \caption{\small Average fixation time $\langle t\rangle$ divided by $k$
    versus $b$ for various initial populations $k$.  The data are based on
    $10^6$ realizations for each point.  The straight line is the prediction
    of Eq.~\eqref{t-kk}.}
\label{T-vs-b}
\end{figure}

There is also an important effect that is not accounted for in
Eq.~\eqref{beta}---the motions of the middle and right particles are
correlated.  When an A either gives birth or dies, the middle and right
particles in Fig.~\ref{fvm} move in lockstep.  This implies that the motion
of the effective particle in the wedge is not isotropic.  While we do not
know how to account for these two effects rigorously---averaging over the
interval and the correlation in the effective particle motions---our
heuristic approach gives the qualitatively correct dependence of the survival
probability exponent $\beta$ on $b$.  We can also get a sense of the role of
correlations in the effective particle motions on the exponent $\beta$ by
considering the $b\to\infty$ limit.  Here, voter model updates do not occur,
so there is no position dependence in the particle diffusion coefficients.
Now Eq.~\eqref{beta} should be directly applicable and it gives
$\beta = \pi/\big[2 \cos^{-1}(1/\sqrt{3})\big]\approx 1.644$, whereas, the
exact exponent from Eq.~\eqref{S2} is 2.

Finally, we investigate the dependence of the fixation time on the birth rate
(Fig.~\ref{T-vs-b}).  The primary observation from these simulations is that
the fixation time is a monotonically decreasing function of the birth rate
$b$.  That is, population fluctuations promote fixation.  In a related vein,
the more stable (less volatile) species is more likely to fixate for non-zero
voting rate.

\section{OUTLOOK}

We investigated basic properties of fixation in a fluctuating population.
The population consists of two distinct species, A and B, that are identical
in all dynamical respects, except for their label.  The population of each
species grows and shrinks by the classic birth-death process and, in
addition, AB pairs can transform to AA or BB by voter model dynamics.

In the limit of voting rate $v=0$, the system reduces to two independent
birth-death processes, for which many interesting results can be derived
analytically.  Although a single birth-death process has an infinite average
extinction time, the fixation time for two independent birth-death processes
(the time when one species first goes extinct) is finite.  The distribution
of fixation times asymptotically decays as $t^{-3}$, while the distribution
of the number $n$ of surviving species decays as $n^{-3}$.  These properties
are robust with respect to the initial condition and also to different
birth/death rates for each species.  When the voting rate is non-zero, a
basic outcome is that the fixation time is reduced by birth-death
fluctuations.  That is, population volatility leads to quicker extinction.

There are many directions for future research.  We only investigated
situations where the birth and death rates for each species are equal, and
extending to unequal birth and death rates may reveal new phenomena.  It
would also be interesting to include spatial degrees of freedom into the
dynamics, as this aspect naturally arises in any bacterial colony, and
fixation phenomena have been extensively investigated in growing bacterial
colonies (see~\cite{KAHN10} for a review).

We thank Jacopo Grilli for helpful discussions.  DB and SR gratefully
acknowledge NSF financial support from grant DMR-1608211.  JP acknowledges
support from ``Mar\'ia de Maezt\'u'' fellowship MDM-2014-0370-17-2, from Botin
Foundation, by Banco Santander through its Santander Universities Global
Division and by FIS2015-67616-P.

\newpage
\appendix

\section{Generating Function Solution of the Birth-Death Process}
\label{sec:gf}

We outline some basic facts about the classic birth-death process, in which a
population of independent organisms grows or shrinks because each organism
gives birth at rate $\lambda$ or dies with rate $\mu$.  Let $P_n(t)$ denote
the probability that there are $n$ organisms at time $t$.  This probability
changes with time by according to
\begin{subequations}
\begin{align}
  \dot P_n = \lambda \big[(n-1)P_{n-1}-nP_n\big]+\mu\big[(n+1)P_{n+1}-nP_n\big]\,.
\end{align}
The relevant case is that of equal birth and death rates, so that the average
population is stationary.  In this limit, the master equation reduces to
\begin{align}
  \label{Pdot}
  \dot P_n = (n-1)P_{n-1}-2nP_n+(n+1)P_{n+1}\,,
\end{align}
\end{subequations}
where we set $\lambda=\mu=1$.

A convenient way to solve these equations is by the generating function
method~\cite{KRB10,W05}.  We define the generating function
$g(z,t) = \sum_{n\geq 0}P_n z^n$, multiply Eq.~\eqref{Pdot} by $z^n$, and sum
over all $n$.  After some standard manipulations that involve converting
terms like $\sum_n nP_n z^n$ into a derivative with respect to $z$, the
generating function satisfies $g_t = (1-z)^2 g_z$, where the subscripts
denote partial differentiation.  We convert this to the elementary wave
equation $g_t=g_y$ by defining the variable $dy=dz/(1-z)^2$, from which we
obtain $y=1/(1-z)$, or $z=1-y^{-1}$.  The solution to the wave equation is
$g(y,t)= F(y+t)$, where $F$ is an arbitrary function that is fixed by the
initial condition.

For the single particle initial condition, $P_n(t\!=\!0)=\delta_{n,1}$.  Then
$g(z,t\!=\!0)=z$.  Because the natural variables for the generating function
are $(y,t)$ instead of $(z,t)$, we re-express the initial generating as
$g(y,t=0)= z = F(y)= 1-y^{-1}$.  Since the generating function depends on the
variable combination $y+t$, we have, for $t>0$, $g(y,t) = 1-(t+y)^{-1}$.
Finally, we re-express the generating function in terms of $(z,t)$ to give
\begin{subequations}
\begin{align}
  g(z,t) = 1-\frac{1}{t+\frac{1}{1-z}}~.
\end{align}
We now write this last expression in a Taylor series in $z$ to extract $P_n$
and $P_0$ given in Eq.~\eqref{Pnt}.  To incorporate an arbitrary birth rate
$b$, as in \eqref{Pnt}, we merely make the substitution $t\to bt$.

The above derivation can be straightforwardly extended to the initial
condition of $k$ particles.  Now the initial generating function is
$g(z,t\!=\!0)= z^k$, and following the steps of the previous paragraph, the
generating function is
\begin{align}
  \label{g-k}
  g(z,t) = \left(1-\frac{1}{t+\frac{1}{1-z}}\right)^k\,.
\end{align}
\end{subequations}
For the two-particle initial condition ($k=2$), the Taylor series expansion
of the generating function give $P_n$ and $P_0$ written in Eq.~\eqref{Pnt2}.
For larger $k$, the Taylor series expansion of $g(z,t)$ becomes progressively
more unwieldy.  However, the form of $P_0$ for general $k$ is simple:
$P_0(t)=\big[t/(1+t)\big]^k\to \big[bt/(1+bt)\big]^k$.

\section{Additional Examples}

The presentation in Sec.~\ref{subsec:2} can be readily extended to other
initial conditions and to more than two uncoupled birth-death processes.
Because these examples have illustrative value, we discuss these two cases
below.

\subsection{The Initial state: (A,B) = (2,1)}

We first generalize the derivations in Sec.~\ref{subsec:2} to unequal initial
numbers of particles of each species.  For specificity, we treat the initial
state of of 2 A's and 1 B; it is straightforward to extend our approach to
more general initial conditions.  For a single birth-death process starting
with two particles, the distribution $P_n(t)$ now is (from the Taylor series
expansion of Eq.~\eqref{g-k})
\begin{align}
  \label{Pnt2}
  P_n(t) = \frac{2(bt)^{n}+(n-1)(bt)^{n-2}}{(1+bt)^{n+2}}\qquad \qquad  P_0(t) =
  \left(\frac{bt}{1+bt}\right)^2\,.  
\end{align}
This distribution satisfies $\sum_{n\ge 0}P_n=1$ and conservation of the
average particle number, $\langle n(t)\rangle= \sum_{n\ge 1}nP_n=2$.  From
this expression for $P_0$, the probability that the A's survive until time
$t$, for the $(2,1)$ initial condition, is
\begin{subequations}
  \label{FSA}
  \begin{align}
  S^{(A)}\!(t)=1-P_0 = \frac{1+2bt}{(1+bt)^2}\,,
\end{align}
from which the probability that A's go extinct at time $t$ is
\begin{align}
F^{(A)}\!(t)= -\frac{d}{dt} S^{(A)}(t)= \frac{2b^2t}{(1+bt)^3}\,,
\end{align}
\end{subequations}
while $S^{(B)}$ and $F^{(B)}$ are again given by~\eqref{SF}.

The probability that both birth-death processes do not go extinct by time $t$
is $S_2(t)=S^{(A)}\!(t)\,S^{(B)}\!(t)$, while the probability that extinction occurs at
time $t$ is (without regard to which species goes extinct)
\begin{align}
F_2(t)=  -\frac{d}{dt}S_2(t) = \frac{b(1+4bt)}{(1+bt)^4}\,.
\end{align}
In analogy with \eqref{tav}, the average time for the first extinction to
occur, irrespective of which species goes extinct is now
\begin{align}
  \langle t\rangle =\int _0^\infty dt\, t\, F_2(t) =\int _0^\infty dt\,
  S_2(t) =\frac{3}{2b}\,.
\end{align}
The extinction time is longer than in Eq.~\eqref{tav} because the population
initially is ``further'' from extinction---three particles rather than two.
It is also natural to ask which of the two species goes extinct first.  The
probability $\mathcal{E}^{(A)}$ that species A goes extinct first is
\begin{align}
  \mathcal{E}^{(A)} = \int_0^\infty dt\, F^{(A)}\!(t) \,S^{(B)}\!(t)
  = \int_0^\infty dt\, \frac{2t}{(1+t)^4}=\frac{1}{3}\,.
\end{align}
In this integral, the factor $F^{(A)}$ ensures that A's go extinct at time
$t$ while the factor $S^{(B)}$ ensures that the B's are not extinct at this
time.  Similarly, the probability $\mathcal{E}^{(B)}$ that species B first
goes extinct equals $\frac{2}{3}$.

Finally, the probability that the population consists of $n$ particles of
type A at the moment of B extinction is
\begin{subequations}
  \label{QAB}
\begin{align}
Q_n^{(A)} = \int_0^\infty dt\, F^{(B)}\!(t)\, P_n^{(A)}\!(t) &=
  \int_0^\infty dt\, b\, \frac{2 (bt)^n+(n\!-\!1)(bt)^{n-2}}{(1+bt)^{n+4}}\nonumber\\[3mm]
  &=\begin{cases}  {\displaystyle \frac{4(n+6)\Gamma(n)}{\Gamma(n+4)}}&
    \qquad n>1\\[4mm]
    {\displaystyle \frac{1}{6}} & \qquad n=1\,.
\end{cases}
\end{align} 
Similarly, the probability that the population consists of $n$ particles of
type B at the moment of A extinction is
\begin{align}
Q_n^{(B)} = \int_0^\infty dt\, F^{(A)}\!(t)\, P_n^{(B)}\!(t) =
  \int_0^\infty dt\,b\,\frac{2(bt)^n}{(1+bt)^{n+4}}
  = \frac{4\Gamma(n+1)}{\Gamma(n+4)}\,.
\end{align}
\end{subequations}
Both of the distributions in Eqs.~\eqref{QAB} asymptotically scale as $4n^{-3}$ for
$n\to\infty$.

The distributions $Q_n$ satisfy the basic sum rules:
\begin{align}
\sum_{n\geq 1}  Q_n^{(A)}=\frac{2}{3}\qquad   \sum_{n\geq 1}  Q_n^{(B)}=\frac{1}{3}\,;\qquad\qquad
\sum_{n\geq 1} n Q_n^{(A)}=2\qquad \sum_{n\geq 1} n Q_n^{(B)}=1\,.
\end{align}
The first two relations state that the probability that A's are the surviving
species equals $\frac{2}{3}$, while B's are the surviving species with
probability $\frac{1}{3}$.  The next two relations state that the average
number of A's, conditioned on B's going extinct, equals 2, while the average
number of B's, conditioned on A's going extinct, equals 1.  Thus the average
number of surviving particles at the moment of extinction, independent of
their identity, equals 3.

\subsection{Two Distinct Birth-Death Processes}

Suppose that the common birth/death rates for the two species are different;
we denote these rates as $a$ and $b$ for species A and B, respectively.  The
probability that both birth-death processes do not go extinct by time $t$ is
(compare with Eq.~\eqref{S1})
\begin{align}
  S_2(t)= \frac{1}{(1+a t)} \frac{1}{(1+b t)}\,,
\end{align}
which we again term the survival probability.  The probability that one of
the species goes extinct at time t is (compare with Eq.~\eqref{F1})
\begin{align}
  \label{F2}
  F_2(t) = -\frac{d S_2(t)}{dt} 
         =   \frac{a}{(1+a t)^2(1+b t)}+\frac{b}{(1+b
           t)^2(1+a t)}= F_1^{(A)}\!(t)\,S_1^{(B)}\!(t)+ F_1^{(B)}\!(t)\,S_1^{(A)}\!(t)\,,
\end{align}
where the superscripts refer to the species type.  The first term on the
right-hand side is the probability that species A goes extinct at time $t$
while species B survives, and vice versa for the second term.  The average
extinction time, independent of which species goes extinct, is
\begin{align}
  \langle t\rangle =\int_0^\infty dt\, t\, F_2(t)= \frac{\ln(b/a)}{b-a}\,.
\end{align}                     
For $a, b$ both approaching the common value $b$, the above result reduces to
$\langle t \rangle={1}/{b}$, given in Eq.~\eqref{tav}.

\begin{figure}[ht]
  \centerline{\subfigure[]{\includegraphics[width=0.43\textwidth]{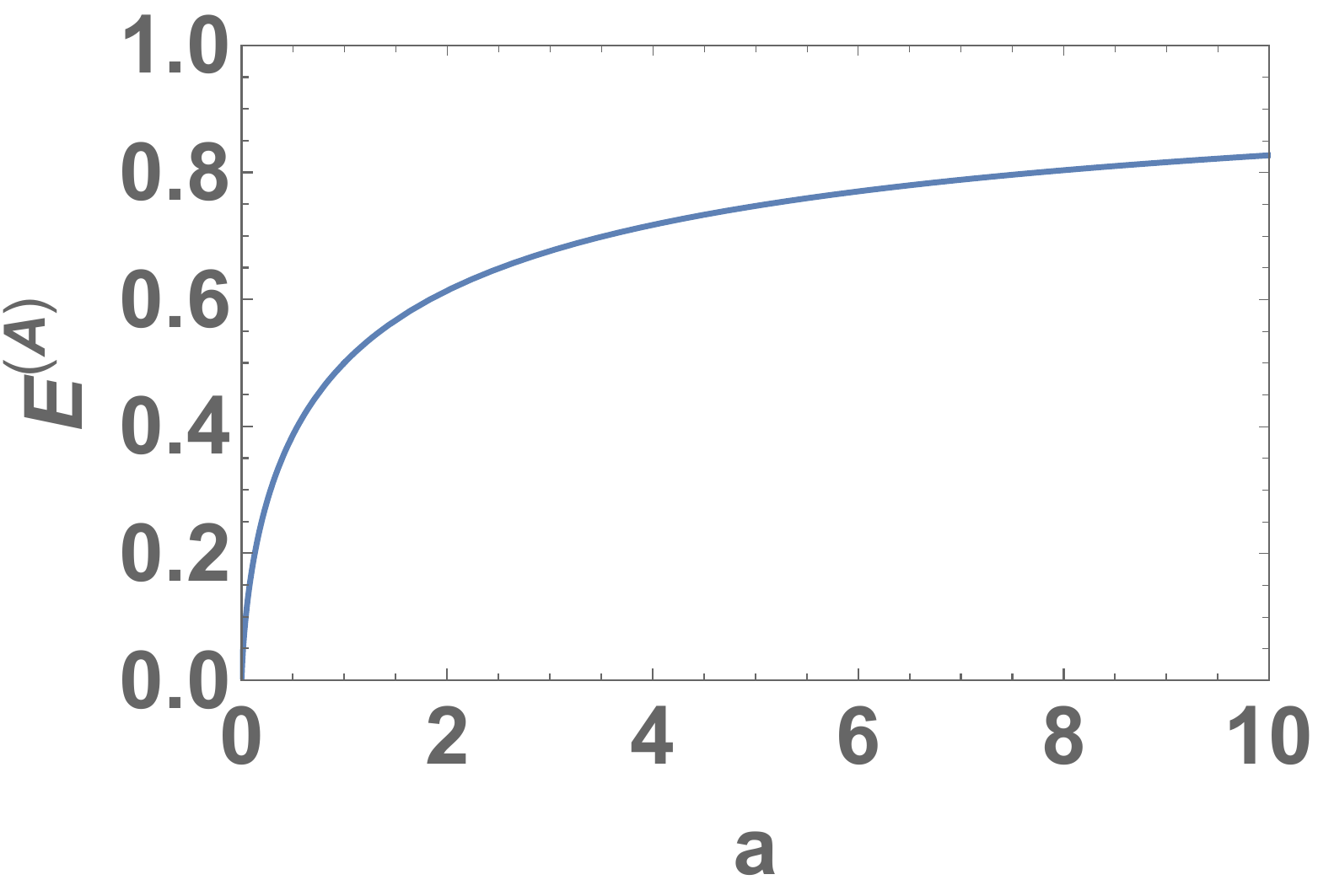}}
  \qquad\qquad \subfigure[]{\includegraphics[width=0.42\textwidth]{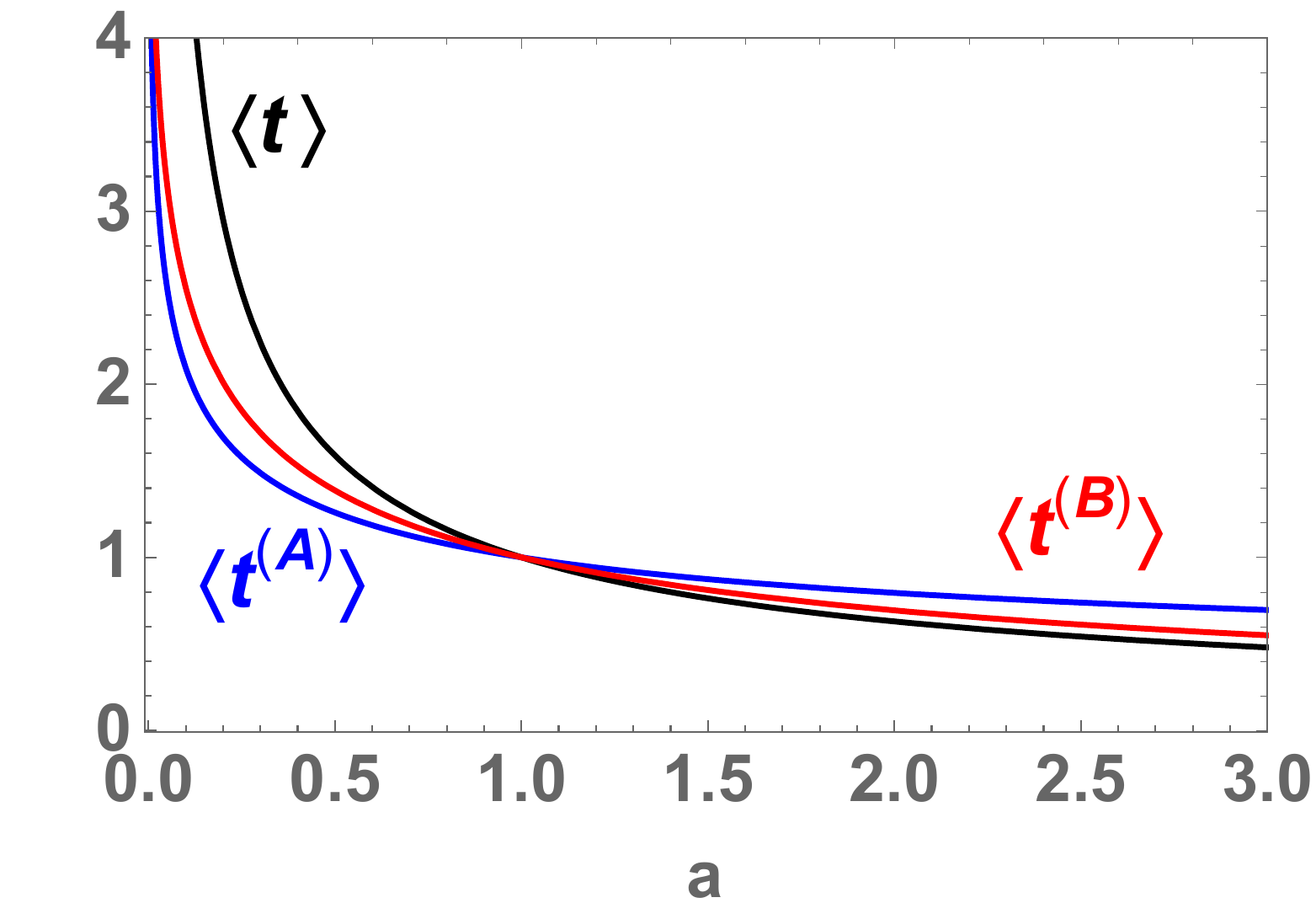}}}
\caption{\small (a) The exit probability $E^{(A)}$ that species A goes
  extinct first as a function of its birth rate $a$.  (b) Unconditional and
  conditional extinction times as a function of the birth rate $a$.  In both
  panels, the birth/death rate of species B is fixed at $b=1$ and the initial
  state contains of one particle of each species. }
\label{E-t}  
\end{figure}

It is natural to ask which species is more likely to go extinct---the more
volatile or the more stable species.  The probability $E^{(A)}$ that species
A goes extinct is
\begin{align}
  \label{Ea}
  E^{(A)} = \int_0^\infty dt\, F_1^{(A)}\!(t)\, S_1^{(B)}\!(t)
= \int_0^\infty dt\, \frac{a}{(1+a t)^2} \frac{1}{(1+b t)}
    = \frac{a\big[a-b+b\ln(b/a)\big]}{(a-b)^2}\,.
\end{align}
The factor $F_1^{(A)}$ ensures that it is species A that goes extinct, while
the factor $S_1^{(B)}$ ensures that B's still survive when A goes extinct.
Integrating this product over all time gives the total probability that
species A goes extinct.  From \eqref{Ea}, it is likelier that the more
volatile species goes extinct for the symmetric initial condition
(Fig.~\ref{E-t}(a)).

We also determine the \emph{conditional} extinction times, namely, the
average time for a specified species to go extinct.  The average time for
species A to go extinct is given by
\begin{align}
  \langle t^{(A)}\rangle = \int_0^\infty dt\, t\, F_1^{(A)}\!(t)\, S_1^{(B)}\!(t)\bigg/
\int_0^\infty dt\,  F_1^{(A)}\!(t) \,S_1^{(B)}\!(t)= \frac{b-a +a\ln
  (b/a)}{a\big[b-a+b\ln(b/a)\big]}\,.
\end{align}
The average time $\langle t^{(B)}\rangle$ for species B to go extinct is just
the above expression with $a$ and $b$ interchanged.  Figure~\ref{E-t}(b)
shows these extinction times for $b=1$ and varying $a$; we see that increased
volatility decreases the extinction time.

\begin{figure}[ht]
  \centerline{\includegraphics[width=0.45\textwidth]{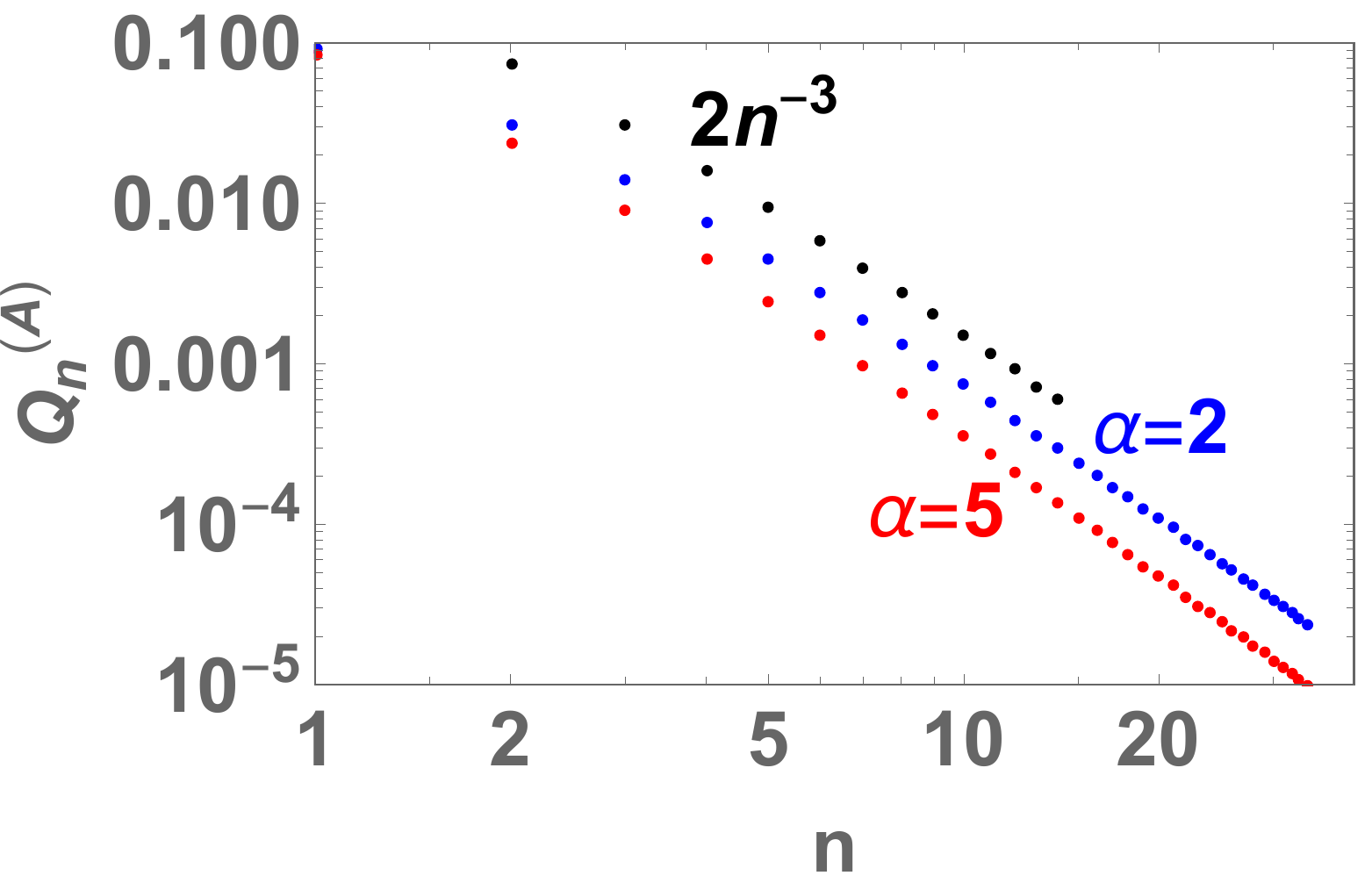}}
  \caption{\small Numerical integration of $Q_n^{(A)}$ in Eq.~\eqref{Qn-asymm}
    versus $n$ on a double logarithmic scale for $n\leq 40$.}
\label{Qn_numerical}  
\end{figure}

At the instant when species B goes extinct, Mathematica~\cite{W} gives the
distribution of the number of species A that remain as (compare with
Eq.~\eqref{Qn})
\begin{align}
\label{Qn-asymm}
  Q_n^{(A)} &= \int_0^\infty dt\,\ F_1^{(B)}\!(t)\, P_n^{(A)}\!(t)
= \int_0^\infty dt\,\,
  \frac{b}{(1+b t)^2}\, \frac{(a t)^{n-1}}{(1+a t)^{n+1}}\nonumber \\
  &=
\frac{\alpha}{(\alpha\!-\!1)^3}
    \left\{\!\!\bigg[\frac{\alpha}{n}+\frac{\left(n^2+1\right)}{n(n-1)}
    -\frac{\big[2\alpha+ (n\!-\!1)\big] \,
    _2F_1\left(1,1;3\!-\!n;\alpha\right)}{n\!-\!2}\bigg]\right. \nonumber\\
  &\left. \hspace{1.4in} -\frac{\pi \alpha^{n-1}}{(1-\alpha)^{n+2}}\big[2\alpha+1 (n\!-\!1)\big] \csc (n\pi)\right\},
\end{align}
where $\alpha =a/b$.  Unfortunately, this representation is pathological for
all positive integer $n$: the hypergeometric function $_2F_1$ diverges for
all $n\geq 3$, so that the first square bracket is diverges for all
$n\geq 0$, but these divergences are all canceled by the term
$\text{csc}(n\pi)$.  A numerical evaluation of this integral clearly shows
that $Q_n$ asymptotically scales as $n^{-3}$ for all $\alpha$, with a
coefficient that is a decreasing function of $\alpha$.

\subsection{$k$ Symmetric Uncoupled Species}

Finally, we treat the case of $k$ distinct species that all have common
birth/death rates.  We treat the initial condition of a single particle of
each species.  As a function of time, a series of partial extinctions occurs,
in which the number of extant species decreases by 1 before the final
extinction where only a single species remains.  To determine the time for
the first extinction, we use the fact that the probability that $k$
independent birth-death processes do not go extinct before time $t$ is
$S_k(t)=[S_1(t)]^k$, with $S_1$ given by \eqref{S1}.  Thus $S_k(t)$ is the
probability that the first extinction time is $t$ or greater.  The
probability that this first extinction occurs at time $t$ therefore is
(compare with Eq.~\eqref{F1})
\begin{align}
  F_k(t)&= -\frac{d S_k(t)}{dt}   = k [S_1(t)]^{k-1} F_1(t) = \frac{ kb}{(1+b t)^{(k+1)}}\,,
\end{align}
and the average time for the first extinction is
\begin{align}
\langle t\rangle =\int_0^\infty dt\, t \, F_k(t)
                  =\int_0^\infty dt\, S_k(t) = \frac{1}{(k-1)b}\,.
\end{align}

The number of particles of each species at the first extinction can be
obtained by particle conservation.  When there is 1 particle of each species
in the initial state, these $k$ initial particles will be equally distributed
among the $k-1$ remaining species at the first extinction.  Thus there will
be $k/(k-1)$ particles of each species, on average, at the first extinction.
At each subsequent extinction, the $k$ initial particles will be equally
distributed among the remaining species.

For the initial state that consists of $k$ distinct species, with one
particle of each species, we also calculate $Q_n^{(k)}$, the distribution of
the number of particles in one of the $k-1$ remaining species at the first
extinction event.  The generalization of Eq.~\eqref{Qn} is
\begin{align}
Q_n^{(k)} &=\frac{k}{k\!-\!1}\int_0^\infty dt \,F_{k-1}(t) P_n(t) 
  =k\int_0^\infty dt \,F_1(t) P_n(t) \big[S_1(t)\big]^{k-2}\nonumber\\
          & =\frac{k\Gamma(k\!+\!1)\Gamma(n)}{\Gamma(k\!+\!n\!+\!1)}
\simeq k\Gamma(k\!+\!1) n^{-(k+1)}\,.
\end{align}
The prefactor $k$ accounts for the fact that any of the $k$ initial species
could go extinct first, while the factor $k-1$ in the denominator arises
because we are counting only one of the $k-1$ remaining species.  With these
definitions, we recover the obvious sum rules, $\sum_{n\geq 1}Q_n^{(k)}=1$
and $\sum_{n\geq 1}nQ_n^{(k)}=k/(k-1)$.  In each subsequent extinction, the
distribution of the number of particles in any one of the remaining species
becomes gradually broader until $Q_n\sim n^{-3}$ when only a single species
remains.

\end{document}